\documentclass[aip,apl,preprint,showpacs,superscriptaddress,floatfix]{revtex4-1}
\usepackage{amsmath}
\usepackage{amssymb}
\usepackage{graphicx,color}
\usepackage{epstopdf}
\usepackage{ulem}
\makeatletter

\newcommand{\Rmnum}[1]{\expandafter\@slowromancap\romannumeral #1@}
\makeatother

\begin{document}

\title{Diamond/c-BN HEMTs for power applications: A theoretical feasibility analysis}

\author{Namita Narendra}
\affiliation{Department of Electrical and Computer Engineering, North Carolina State University, Raleigh, NC 27695, USA}

\author{Jagdish Narayan}
\affiliation{Department of Materials Science and Engineering, North Carolina State University, Raleigh, NC 27695, USA}

\author{Ki Wook Kim}\email{kwk@ncsu.edu}
\affiliation{Department of Electrical and Computer Engineering, North Carolina State University, Raleigh, NC 27695, USA}
\affiliation{Department of Physics, North Carolina State University, Raleigh, NC 27695, USA}

\begin{abstract}
Diamond is a promising material for high-power electronic applications in both the dc and rf domains.  However, the predicted advantages are yet to be realized for a number of technical challenges. In particular, n-type devices have not been feasible due to the large ionization energies and low thermodynamic solubility limits of n-dopants. Motivated by the recent advances in nonequilibrium processing, we propose and theoretically examine a diamond/c-BN HEMT that can circumvent the critical limitations.  A first-principles calculation suggests the desired type-I alignment at the heterojunction of these two nearly lattice matched semiconductors.  The investigation also illustrates that a large sheet carrier density in excess of $5\times10^{12} ~\mathrm{cm^{-2}}$ can be induced in the undoped diamond channel by the gate bias.  A subsequent analysis of a simple prototype design indicates that the proposed device can achieve large current drive ($ \sim 10 ~ \mathrm{A/cm} $), low $R_{on}$  ($\sim 0.05 ~ \mathrm{m}\Omega \cdot \mathrm{cm}^2$), and high $f_T$ ($ \sim 300 ~\mathrm{GHz}$) simultaneously.
\end{abstract}

\maketitle

The current state-of-the-art power devices based on 4H-SiC and GaN have certain limitations due to their intrinsic material properties.~\cite{Ozpineci2003} Diamond is a promising alternative as evidenced by the favorable figures of merit.~\cite{Balmer2008} This is a result of the superior material properties exhibited by diamond such as band gap (5.45 eV), critical field ($\sim$10 MV/cm), mobility (4500 cm$^2$/V$\cdot$s for electrons and 3800 cm$^2$/V$\cdot$s for holes) and thermal conductivity (20 W/cm$\cdot$K).~\cite{Balmer2008} The theoretical predictions also indicate that the lower limit of the specific on-resistance ($R_{on}$) is about two orders of magnitude smaller for diamond than GaN as can seen from Fig.~1. The resulting improvements in device size and power density make diamond ideal for high-power, high-frequency, and high-temperature applications.
But, the promise of diamond power devices has not been fulfilled yet. Experimentally there have been material challenges such as the lack of high-quality, large-area crystalline diamond films. In addition, the large activation energy for n-type dopants (e.g., $\sim$0.6 eV for phosphorous)~\cite{Kanda2000} has restricted the available diamond power devices to only p-type~\cite{Kohn2005} due to the negligible free electron density.

However, recent advances in non-equilibrium growth and doping control may overcome the technical challenges that have long plagued this ultra-wide bandgap semiconductor.  It was found that amorphous carbon films can be converted directly into diamond at ambient temperatures and pressures through nanosecond laser melting in a super undercooled state and subsequent rapid quenching.~\cite{Narayan2015}
By providing an epitaxial template such as sapphire or silicon, diamond grows from the liquid phase into a large-area single-crystal film via domain matching epitaxy. Furthermore, these films can be doped with p- and n-type dopants through alloying before the melting stage. During the rapid quenching process from the liquid phase, dopant concentrations can far exceed thermodynamic solubility limits through the solute trapping phenomenon.~\cite{Narayan2016a}  Once the single-crystal film is formed, it is also possible to stack additional layers of epitaxial diamond and/or cubic boron nitride (c-BN, a closely lattice match material with a gap of 6.4 eV) through pulsed laser ablation with selected dopant types and concentrations.  Accordingly, fabrication of multi-layer vertical homo- and hetero-junction structures is clearly within reach.  Indeed, the synthesis of undoped diamond on c-BN has already been achieved.~\cite{Narayan2016b}

These developments open up a particularly interesting prospect of modulation doping that can circumvent the issue of large donor activation energy and subsequent carrier mobility degradation, particularly in n-type devices.~\cite{Koizumi2008} If a type-I band offset can be formed at the diamond/c-BN interface, it is possible to dope the c-BN barrier while the undoped diamond layer can function as an efficient carrier channel with both high carrier density and mobility (i.e., HEMTs). The aim of the present study is to examine, via a theoretical analysis, the feasibility of the diamond/c-BN based HEMTs and gauge their potential performance.

As it is well known, the 2D electron gas formed in a HEMT is highly dependent on the band alignment at the heterointerface. So far, theoretical calculations in the diamond/c-BN system have predicted two widely different values of 1.42 eV~\cite{Pickett1988} and 0.71 eV~\cite{Sakaguchi1998} for the valence band offset ($\Delta E_V$) for the (100) interface, while no experimental measurement is available.  Hence, we estimate the corresponding values for the (111) oriented structure (grown on, for instance, C-sapphire) based on the first-principles density functional theory (DFT) approach.  Specifically, the DFT calculations are carried out using norm conserving pseudopotentials with the generalized gradient approximation of Perdew-Zunger, as implemented in the plane wave based QUANTUM ESPRESSO package.~\cite{Giannozzi2009} The (111) diamond/c-BN supercell is constructed with 6 layers of C and 6 layers of BN, and by setting the lattice constant of the supercell to the optimized lattice constant of diamond (3.54 {\AA}). The supercell is fully relaxed until the total energy is minimized and the total force on the atoms is less than 10$^{-4}$ Ry. In addition, of the various interface atomic configurations possible, the one leading to minimal energy is selected for which $\Delta E_V$ is calculated as~\cite{Baroni1991}
\begin{equation}
\Delta E_V=(E_V^\mathrm{Diamond}-E_V^\mathrm{c-BN})_\mathrm{bulk}+\Delta V .
\end{equation}
The conduction band offset $\Delta E_C$ can then be evaluated by subtracting $\Delta E_V$ from the experimental band gap values. The first two terms in Eq.~(1) refers to the difference in the valence band maximum energy between diamond and c-BN obtained from their respective bulk unit cells. This value is calculated to be 1.4 eV. The third term, $\Delta V$ accounts for the difference in the reference energy in the bulk calculation in addition to the dipolar correction at the interface. It is obtained by calculating the macroscopic average potential across the diamond/c-BN supercell as shown in Fig.~2. After careful periodic averaging of the potential, $\Delta V$ is estimated to be approximately $-0.9$ eV to $-1$ eV. Thus, both $\Delta E_V$ and $\Delta E_C$ arrive at a similar value of around $0.4 - 0.5$ eV with a type-I alignment of the heterojunction. In the rest of the analysis, a conservative value of $\Delta E_C = 0.4$ eV is adopted for device characterization.

Figure~3 shows the schematic of the prototype diamond/c-BN n-HEMT structure under investigation. In addition to the substrate (i.e., the epitaxial template mentioned above), the structure consists of n-diamond layer, undoped diamond channel, undoped c-BN spacer, and n-doped c-BN barrier layer.  The doped c-BN barrier can be followed by an undoped dielectric to prevent surface conduction or other potential degradation.  In c-BN, a doping level of 10$^{18}$ cm$^{-3}$ is assumed with an activation energy of 0.24 eV.  The latter corresponds to the generally accepted ionization energy of Si dopants although a much smaller value of 0.05 eV is also reported in the literature.~\cite{Zhang2013}  Another option is carbon, whose activation energy was measured to be around 0.2 eV at a high dose.~\cite{Narayan2016b} The details of doping in the diamond layer is unimportant due to the low ionization efficiency (e.g., phosphorous with 0.6 eV). As for the gate, a Schottky contact is considered with a barrier height of 0.5 eV while the source and drain electrodes are assumed to be ideal ohmic contacts. Room temperature is considered throughout the study.  The conduction band profile is obtained by using a Poisson simulator that takes into account the incomplete ionization model [Fig.~4(a)]. The resulting electron density distribution clearly shows the formation of a 2D electron channel in the diamond layer with the carrier concentrations exceeding 10$^{18}$ cm$^{-3}$ (at no gate bias). The modulation of the sheet charge density ($n_s$) with a gate voltage ($V_{gs}$) is illustrated in Fig.~4(b) with the drain voltage ($V_{ds}$) set to 0 V. The maximum sheet carrier concentration obtained is $5\times10^{12}$ cm$^{-2}$, which can be optimized further by varying the device parameters.

With the proof of concept for the diamond/c-BN HEMT established, the dc current characteristics are now determined by adopting an analytical charge control model.~\cite{Lee1983}  The HEMT operation can be divided into two regimes; i.e., the linear and the velocity saturation regimes. In the linear regime, assuming total depletion in the c-BN barrier region [as the donor levels are located higher than the diamond conduction band edge as shown in Fig.~4(a)], the solution of the 1D Poisson's equation yields the sheet carrier density as
\begin{equation}
n_s(x)=\frac{\epsilon_\mathrm{c-BN}}{q(d+d_\mathrm{2D})}[V_{gs}-V_{th}-V_{ch}(x)] ,
\end{equation}
where $\epsilon_\mathrm{c-BN}$ is the permittivity of c-BN, $d$ the total thickness of the c-BN region, $d_\mathrm{2D}$ the 2D electron gas thickness, $V_{th}$ the calculated threshold voltage, and $V_{ch}$ the channel voltage resulting from the applied $V_{ds}$. The drain-source current, $I_{ds}$ can then be obtained from the current density equation as
\begin{equation}
I_{ds}=wq\frac{\mu_0\frac{dV_{ch}(x)}{dx}}{1+\frac{1}{E_{cr}}\frac{dV_{ch}(x)}{dx}}n_s(x) .
\end{equation}
Here, $w$ is the gate width, $q$ the unit charge, $\mu_0$ the low-field electron mobility (4500 cm$^2$/V$\cdot$s), and $E_{cr}$  the critical field for velocity saturation (10 kV/cm) in diamond.~\cite{Balmer2008}  As given, the model accounts for mobility degradation at high fields. The final expression for $I_{ds}$ can be obtained by substituting Eq.~(2) in Eq.~(3) and then integrating Eq.~(3) along the channel length by applying appropriate boundary conditions for the channel potential, in which the source and drain parasitic resistances can be taken into consideration. In the saturation regime, the channel can be divided into two regions, a low-field region and a high-field region near the drain end of the channel where the channel is pinched off. Using Eq.~(3), the low-field channel length and the electric field in the corresponding region can be obtained. In the high-field region, the solution to 2D Poisson's equation yields an expression for the channel potential, $V_{ch}(x)$. $I_{ds}$ is obtained by enforcing the current and field continuity between the two regions. The final expressions for $I_{ds}$ can be found in Ref.~\onlinecite{Rashmi2001}.

The simulated $I-V$ characteristics are provided in Fig.~4(c) for different gate biases. A maximum $I_{ds}$ over 10 A/cm can be obtained which is comparable to GaN/AlGaN HEMTs.  $R_{on}$ extracted from the slope of the $I_{ds}$-$V_{ds}$ curve in the linear region indicates a very low value (0.05 $\mathrm{m}\Omega \cdot \mathrm{cm^2}$) in concert with the estimate shown in Fig.~1. The calculated peak transconductance $g_m$ is also large at 580 mS/mm. Both $R_{on}$ and $g_m$ are evaluated for $V_{gs}=2$~V. Further, the unity gain cutoff frequency ($f_T$) is estimated by using parasitic gate capacitances as $f_T=\frac{g_m}{2\pi (C_{gs}+C_{gd})}$.  Here, $C_{gs}$ and $C_{gd}$ are the gate-to-source and gate-to-drain capacitances, respectively, derived from the total charge $Q_\mathrm{sat}$ in the channel in the velocity saturation regime as $C_{g(d)s}=\frac{\partial Q_\mathrm{sat}}{\partial V_{g(d)s}}$. The result suggests $f_T$ nearly in the sub-mm wave range (280 GHz).

In summary, a diamond/c-BN HEMT structure is proposed and evaluated.  The analysis clearly predicts a highly promising performance in key aspects including $I_{ds}$, $R_{on}$, $g_m$, and $f_T$.  With further optimization, it is anticipated that this device can provide a major advance in the high-power, high-frequency, high-temperature electronic applications.

\clearpage

\begin{figure}
\begin{center}
\includegraphics[width=6cm]{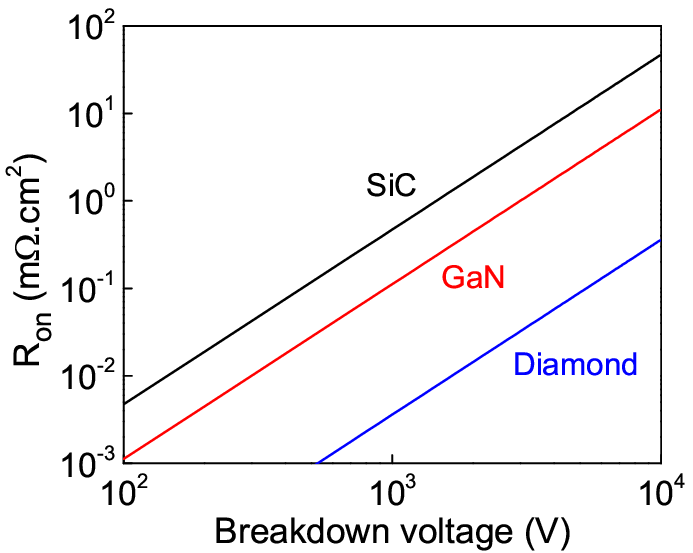}
\caption{Theoretical specific on-resistance versus breakdown voltage for diamond in comparison with 4H-SiC and GaN.}
\end{center}
\end{figure}

\clearpage
\begin{figure}
\begin{center}
\includegraphics[width=5.5cm]{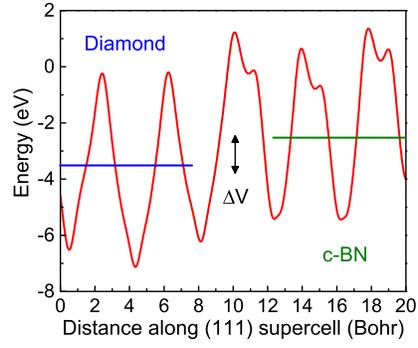}
\caption{Planar averaged potential across the (111) diamond/c-BN heterostructure obtained from a DFT calculation. }
\end{center}
\end{figure}

\clearpage
\begin{figure}
\begin{center}
\includegraphics[width=6cm]{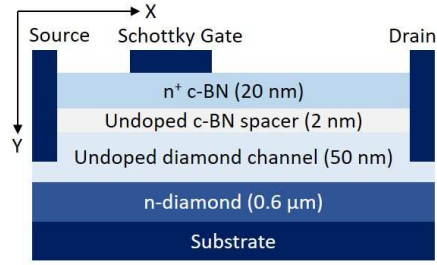}
\caption{Schematic of the proposed diamond/c-BN HEMT prototype. The gate length is 1 $\mu$m, while the gate-to-source and gate-to-drain lengths are 1 $\mu$m and 2 $\mu$m, respectively.}
\end{center}
\end{figure}

\clearpage
\begin{figure*}
\begin{center}
\includegraphics[width=17cm]{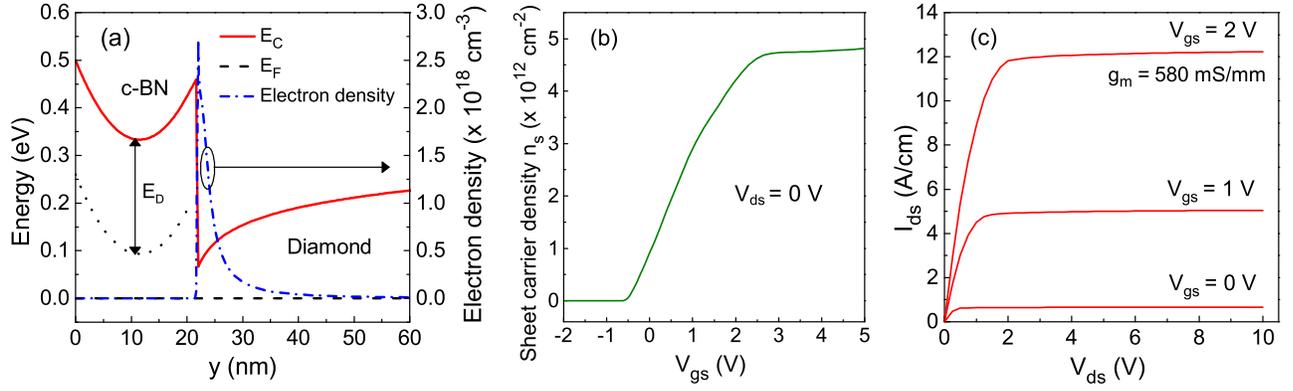}
\caption{(a) Calculated conduction band profile and the corresponding electron density in the channel with no gate bias. The activation energy ($E_D$) for the n-type dopants is also indicated in the c-BN layer. (b) Sheet carrier density as a function of $V_{gs}$ with $V_{ds}$ set to 0 V. (c) Output $I_{ds}-V_{ds}$ characteristics obtained at different $V_{gs}$.  The transconductance $g_m$ is estimated at $V_{gs} =2$ V.}
\end{center}
\end{figure*}

\end{document}